\documentclass[a4paper]{jpconf}
\usepackage{graphicx}
\begin{document}

\title{Non-equilibrium dilepton production in hadronic transport approaches}

\author{Jan Staudenmaier$^{1,2}$, Janus Weil$^1$, Hannah Petersen$^{1,2,3}$}

\address{$ˆ1$ Frankfurt Institute for Advanced Studies (FIAS), Ruth-Moufang-Stra{\ss}e 1, 60438 Frankfurt am Main}
\address{$ˆ2$ Institut f\"ur Theoretische Physik, Johann Wolfgang Goethe-Universit\"at, Max-von-Laue-Str. 1, 60438 Frankfurt am Main, Germany}
\address{$ˆ3$ GSI Helmholtzzentrum f\"ur Schwerionenforschung, Planckstr. 1, 64291 Darmstadt, Germany}

\ead{staudenmaier@fias.uni-frankfurt.de}

\begin{abstract}
In this work the non-equilibrium dilepton production from a hadronic transport approach (SMASH) is presented. The dilepton emission from the hadronic stage is of interest for current HADES results measured at GSI in the beam energy range from 1.25 - 3.5 GeV. Also at high collision energies
(RHIC/LHC) the later dilute stages of the reaction are dominated by hadronic dynamics.
The newly developed hadronic transport approach called SMASH (=Simulating Many Accelerated Strongly-interacting Hadrons) is introduced first. After explaining the basic interaction mechanisms, a comparison of elementary cross sections for pion production to experimental data is shown.
The dilepton production within SMASH is explained in detail. The main contribution to the dilepton spectra in the low energy regime of GSI/FAIR/RHIC-BES originates from resonance decays. Results of the dilepton production with SMASH such as invariant mass spectra are shown.

\end{abstract}

\section{Introduction}

Lepton pairs are clean probes for strongly-interacting matter, since they only interact electromagnetically. They are directly emitted from the hot and dense medium created in a heavy-ion collision, whereas hadronic probes are rescattered or absorbed. It is therefore possible to extract medium properties and medium modifications of resonances from dileptons over the whole lifetime of such collisions. Dileptons are measured for low beam energies by the HADES collaboration at GSI~\cite{HADES:2011ab} and for high beam energies by STAR and PHENIX at RHIC~\cite{Adamczyk:2015lme,Adare:2015ila} or ALICE at LHC~\cite{Koehler:2014dba}.

At the moment it is not possible to describe heavy-ion collisions from first principles, since descriptions of highly-dynamical many-body systems directly based on QCD are not feasible. Therefore, often a combination of effective approaches is used, where the early stage is described by models based on fluctuating color fields or strings, the hot and dense stage is governed by relativistic dissipative hydrodynamics with an equation of state from QCD lattice calculations and the late, dilute stage is described by hadronic transport approaches. Hadronic transport approaches do not rely on the assumption of equilibrium and are able to describe low beam energy collisions. For this work a new hadronic transport approach, SMASH ("Simulating Many Accelerated Strongly-Interacting Hadrons"), is employed that constitutes an effective numerical solution of the relativistic Boltzmann equation.

\section{Model Description}

SMASH~\cite{Weil:2016zrk,Weil:2016fxr} has been developed with the intention to provide a standard reference for the description of a hadronic system with vacuum properties. A geometric collision criterion is employed, where two particles interact, if the distance of closest approach is smaller than a so called {\it interaction distance} that depends on the total cross section. This ansatz is known from other transport models~\cite{Bass:1998ca}. Low-energy nucleus-nucleus collisions, as well as infinite matter calculations can be performed with the approach. It is also possible to set it up as an afterburner for hydrodynamic simulation.

All well-established hadrons from the PDG~\cite{Agashe:2014kda} up to mass of $2 \,\rm{GeV}$ are included in the model. The $\pi,\eta,\rho, \omega, \phi, f_2, \sigma, K$ mesons and the $N,\Delta,\Lambda, \Sigma, \Xi,\Omega$ baryonic states are incorporated.

\begin{figure}[h]
\includegraphics[width=20pc]{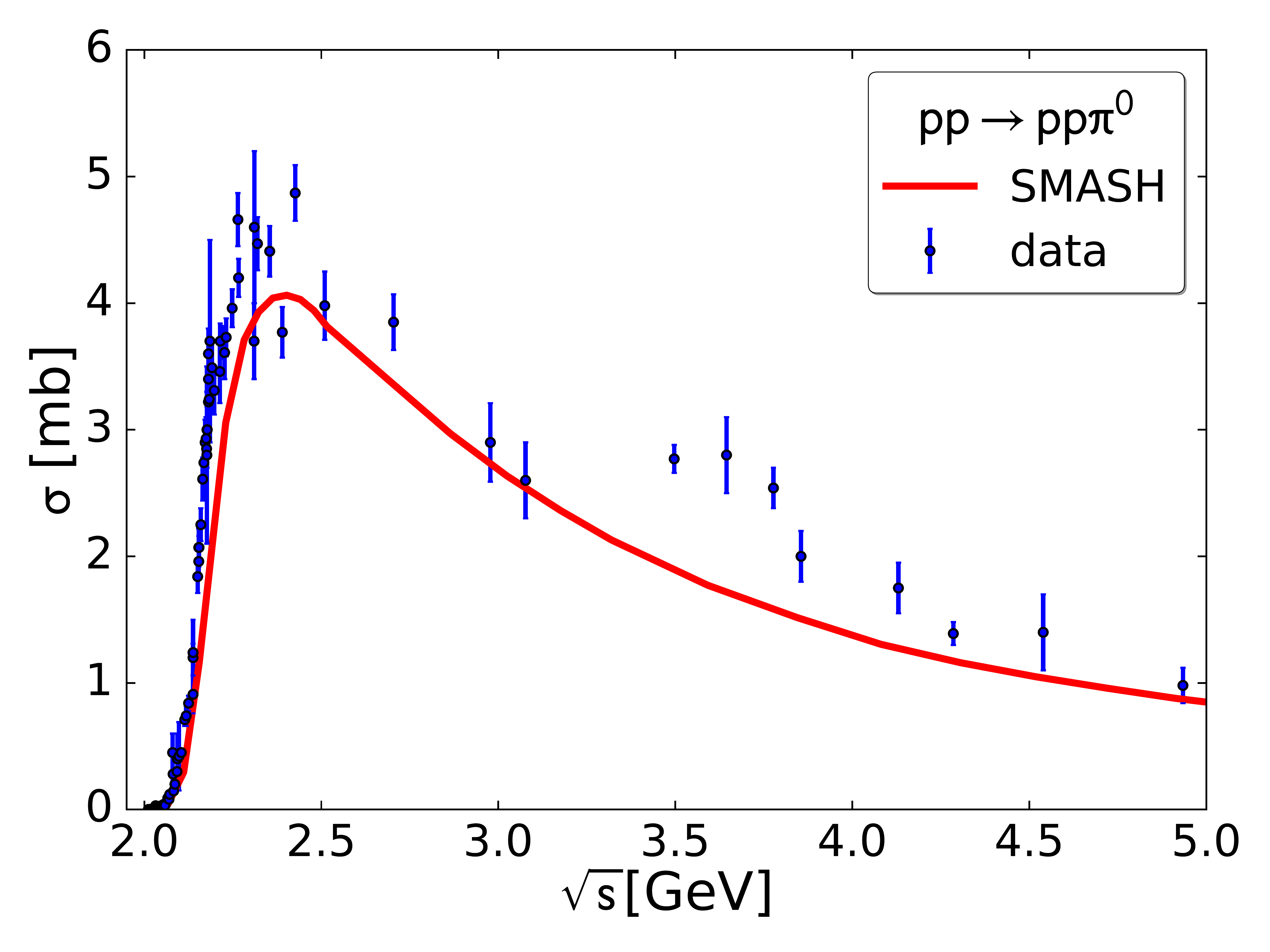}\hspace{2pc}
\begin{minipage}[b]{15pc}\caption{\label{pi_xs} Cross section of the process $pp\rightarrow pp\pi^0$ for different $\sqrt{s}$ energies. Compared to data taken from \cite{LaBoer}. \vspace{2pc}}
\end{minipage}
\end{figure}

In the few GeV energy regime the excitation and decay of resonances dominate the hadronic cross section. Therefore, all interactions fall in one of 4 categories: elastic scatterings, inelastic scatterings, decays and resonance absorption. In order to conserve detailed balance only 1 $\leftrightarrow$ 2 or 2 $\leftrightarrow$ 2 processes are allowed.

Figure~\ref{pi_xs} shows an example for hadronic cross sections. The cross section for the production of one $\pi^0$ in a proton-proton collision is shown for different $\sqrt{s}$. The dominant contribution to this cross section is the excitation of a nucleon state to a $\Delta$ resonance. Only in the higher energy tail additional contributions from other $\Delta^*$ and $N^*$ states contribute. Overall the $\pi^0$ production agrees well with the experimental data. However, a slight systematic undershoot can be observed, which might be due to the negligence of non-resonant contributions to pion production.

\begin{figure}[h]
\includegraphics[width=20pc]{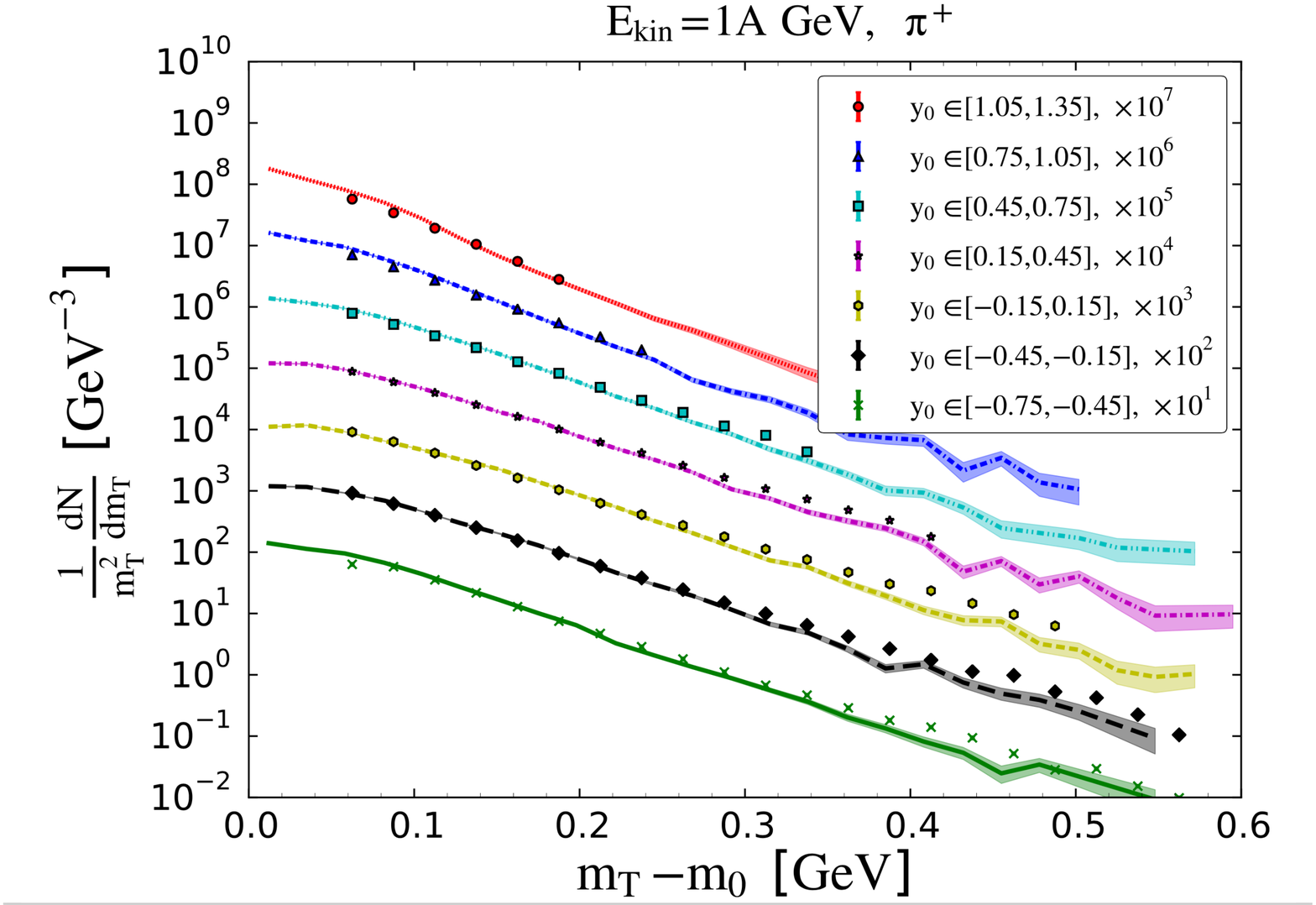}\hspace{2pc}
\begin{minipage}[b]{15pc}\caption{\label{mt}$\pi^+$ $m_T$-spectra in different rapidity windows  for Carbon-Carbon collisions at $E_{\rm{kin}}=1 A\,\textrm{GeV}$ compared with HADES data \cite{Agakishiev:2009zv}. \vspace{2pc}}
\end{minipage}
\end{figure}

In Figure~\ref{mt} a result for a heavy-ion collision at low energies is displayed. The figure shows the $m_T$ spectrum for $\pi^+$ in different rapidity windows for a CC collision at a kinetic energy of $1\,A\rm{GeV}$. Although at high $m_T$ the spectrum is slightly below the data , there is overall a good agreement with the experimental data.

\section{Dilepton Production in SMASH}

Pairs of electrons or muons that are produced together are called dileptons. In SMASH all dileptons are produced through decays. Although di-muons can be produced, in this work  the focus is only on di-electrons. Those originate either from direct decays of the vector mesons $\rho$, $\omega$ and $\phi$ or from Dalitz decays from the pseudoscalar mesons $\pi^0$ and $\eta$ or the $\omega$ and $\Delta$ resonances.

In experiments as well as in theory a challenge for dileptons as an observable is that their decays are very rare. Because these are electromagnetic decays, typical branching ratios are of the order of $10^{-5}$. Therefore in transport models like SMASH the so called {\it Time-Integration-Method} or also called {\it Shining-Method}~\cite{Heinz:1991fn,Schmidt:2008hm, Weil:2012ji} is used. The idea is to integrate the decay probability for the dilepton decay over the whole lifetime of the decaying resonance. In a timestep-based transport approach this can be achieved by radiating ({\it shining}) dileptons during every time step and weighting those pairs according to their decay probability. In this way much more lepton pairs are produced and statistics are improved.

\begin{figure}[h]
\includegraphics[width=20pc]{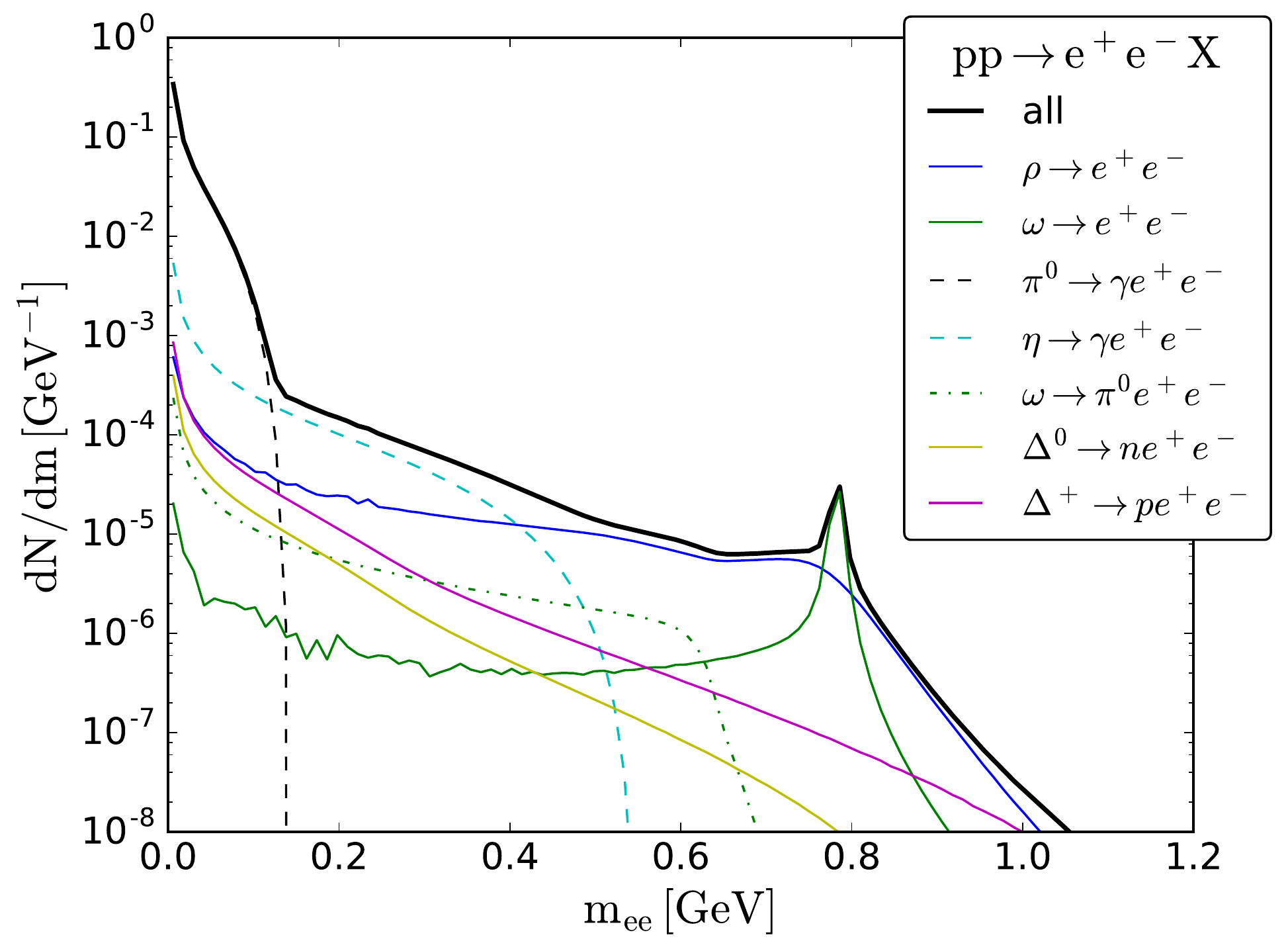}\hspace{2pc}
\begin{minipage}[b]{15pc}\caption{\label{mass}Invariant mass spectrum of di-electrons produced by pp collisions at $E_{\rm{kin}}=3.5\,\textrm{GeV}$.\vspace{2pc}}
\end{minipage}
\end{figure}

The invariant mass spectrum of the dileptons consists of all the different decay channel contributions. Figure~\ref{mass} shows the invariant mass spectrum of di-electrons produced by a proton-proton collision with a kinetic energy of $3.5\,\rm{GeV}$. The spectrum is missing a contribution from the $\phi$ decay, since SMASH currently lacks a production mechanism for $\phi$ mesons in pp collisions. The dominant contribution for low invariant masses is the pion Dalitz decay, whereas in the intermediate mass region the direct decays of vector mesons are dominant. Those contributions also reveal features of the spectral function of the vector mesons. Both, $\omega$ and $\rho$ contributions peak at the resonance pole masses and the $\omega$ peak is sharper, because the resonance has a smaller width. Overall the spectrum looks qualitatively as expected.

\begin{figure}[h]
\includegraphics[width=18pc]{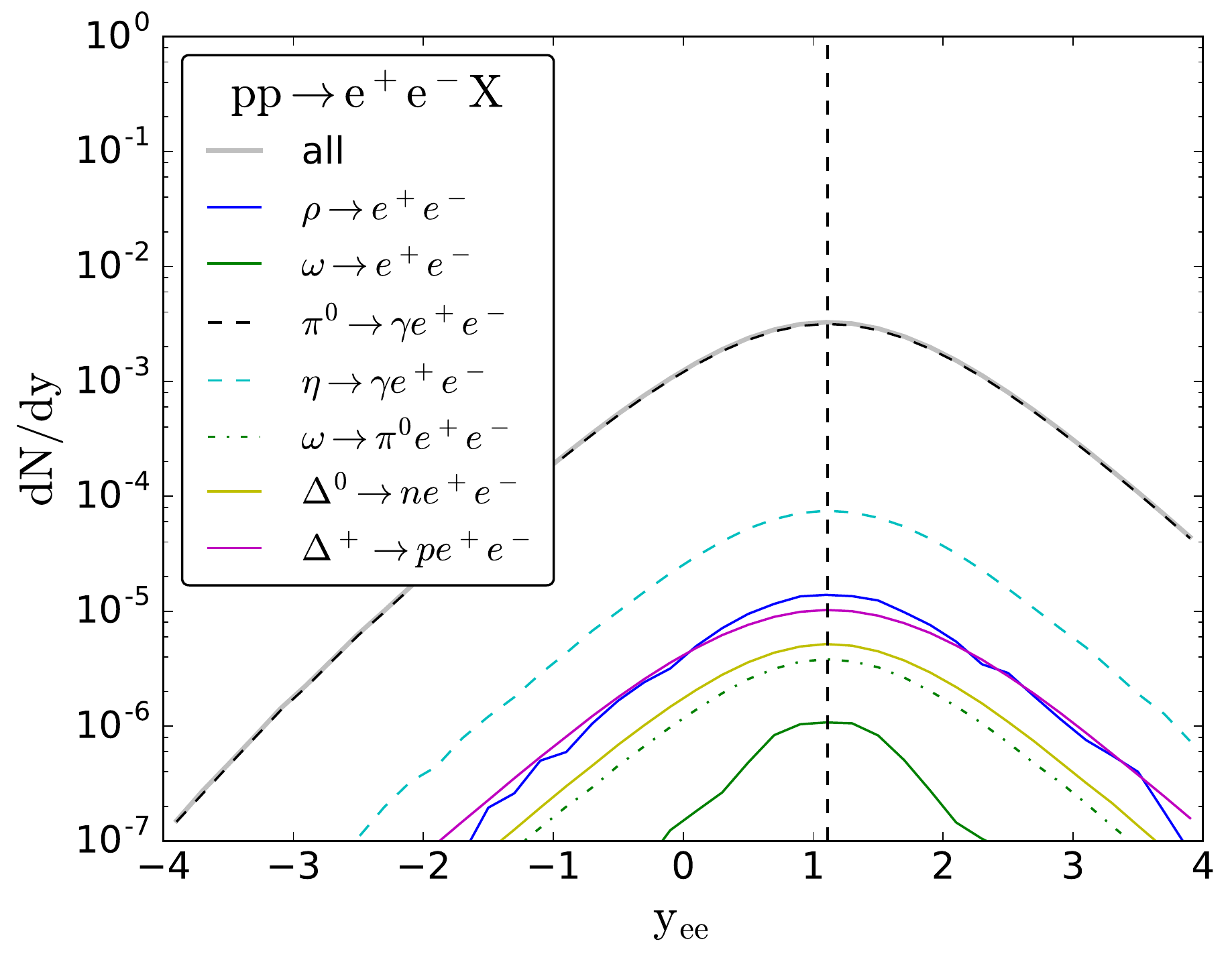}\hspace{4pc}
\begin{minipage}[b]{15pc}\caption{\label{rapidity}Rapidity spectrum of di-electrons produced by pp collisions at $E_{\rm{kin}}=3.5\,\textrm{GeV}$. Vertical dashed line is at mid-rapidity of the pp system in a fixed target setup.\vspace{2pc}}
\end{minipage}
\end{figure}

Since the rapidity spectrum in Figure~\ref{rapidity} does not have a selection in invariant mass, it is clearly visible that the $\pi$ decay is the biggest source of dileptons. Figure~\ref{rapidity} shows the rapidity spectrum of di-electrons of proton-proton collision at $E_{\rm{kin}}=3.5\,\textrm{GeV}$. The whole spectrum is shifted to a finite rapidity, since a fixed-target setup was used. The vertical dashed line is at the mid-rapidity value of the pp system before the collision. Every spectrum peaks at this value, which can be regarded as a check that the kinematic evolution works in a reasonable way.

\section{Summary and Outlook}

One can conclude that SMASH reproduces the hadronic cross sections well and agrees with experimental data on pion production. An extensive description and further comparison with data can be found in~\cite{Weil:2016zrk}. The dilepton production within the approach results in spectra matching the expectations qualitatively. The next step is to validate the dilepton yields by comparing with results for low-energy proton-proton and nucleus-nucleus collision measured by HADES.

Another interesting possibility left for the future is to employ the non-equilibrium dilepton production from a hadronic transport approach combined with a hydrodynamic approach (hybrid model) to explore the behaviour of electromagnetic radiation in hydrodynamics relative to transport approaches.

\ack{J.S. and H.P. acknowledge funding of a Helmholtz Young Investigator Group VH-NG-822 from the Helmholtz Association and GSI. This work was supported by the Helmholtz International Center for the Facility for Antiproton and Ion Research (HIC for FAIR) within the framework of the Landes-Offensive zur Entwicklung Wissenschaftlich- Oekonomischer Exzellenz (LOEWE) program launched by the State of Hesse. Computational resources have been provided by the Center for Scientific Computing (CSC) at the Goethe- University of Frankfurt.}

\section*{References}
\bibliography{hq}

\end{document}